# Observation of strong electron pairing on bands without Fermi surfaces in LiFe$_{1-x}$Co$_x$As

H. Miao[1], T. Qian[1,*], X. Shi[1], P. Richard[1,2], T. K. Kim[3], M. Hoesch[3], L. Y. Xing[1], X.–C. Wang[1], C. –Q. Jin[1,2], J. –P. Hu[1,2,4] and H. Ding[1,2,*]

[1]Beijing National Laboratory for Condensed Matter Physics, and Institute of Physics, Chinese Academy of Sciences, Beijing 100190, China

[2]Collaborative Innovation Center of Quantum Matter, Beijing, China

[3]Diamond Light Source, Harwell Campus, Didcot, OX11 0DE, United Kingdom

[4]Department of Physics, Purdue University, West Lafayette, Indiana 47907, USA

**In conventional BCS superconductors, the quantum condensation of superconducting electron pairs is understood as a Fermi surface (FS) instability, in which the low-energy electrons are paired by attractive interactions. Whether this explanation is still valid in high-$T_c$ superconductors such as cuprates and iron-based superconductors remains an open question[1,2,3]. In particular, a fundamentally different picture of the electron pairs, which are believed to be formed locally by repulsive interactions, may prevail[4-10]. Here we report a high-resolution angle-resolved photoemission spectroscopy study on LiFe$_{1-x}$Co$_x$As. We reveal a large and robust superconducting (SC) gap on a band sinking below the Fermi energy upon Co substitution. The observed FS-free SC**

**order is also the largest over the momentum space, which rules out a proximity effect origin and indicates that the SC order parameter is not tied to the FS as a result of a FS instability.**

Two main categories of theoretical descriptions arise when trying to describe the high-$T_c$ superconductivity of the iron-based superconductors (IBSCs): the weak coupling approach, which involves only the low-energy electronic structure near the Fermi energy ($E_F$)[11-14], and the strong coupling approach, which emphasizes the local magnetic moments and strong Coulomb interactions[6-10]. In the former, superconductivity emerges as a Fermi surface (FS) instability and is in principle sensitive to FS changes. In particular, the superconducting (SC) gap is tied to the FS and its amplitude is strongly influenced by the nesting conditions. In the latter, the pairing is caused by local antiferromagnetic exchange couplings, well defined in the real space, which lead to a SC order parameter (OP) that is fixed in the momentum space and relatively insensitive to small changes of the electronic structure near the FS. In principle, one can distinguish between these two approaches and get critical information on the pairing mechanism of IBSCs by tracking precisely the evolution of the SC OP on bands for which the contributions to the FS vary drastically. In this respect, LiFe$_{1-x}$Co$_x$As offers a perfect platform for this study because it undergoes a Lifshitz transition with one FS disappearing at small Co substitution[15,16].

We first look at the FS topologies of pristine LiFeAs and LiFe$_{0.97}$Co$_{0.03}$As, which are illustrated in Fig. 1. In agreement with previous studies, the substitution of Co introduces electron carriers and effectively moves the chemical potential upward[15-18]. Since the α FS shown in Fig. 1a only barely crosses $E_F$ in pristine LiFeAs[19,20], a slight substitution of Fe by Co removes this tiny FS pocket at the Γ(0,0) point, and thus the system undergoes a Lifshitz transition[16]. The remnant intensity at $E_F$ around Γ in Fig. 1b is attributed to the limited energy resolution setting (~ 14 meV) for this normal state (NS) measurement (T = 30 K), which broadens the spectral width beyond $E_F$.

To accurately determine the band top of the α band, we performed high-resolution (~ 3 meV) angle-resolved photoemission spectroscopy (ARPES) measurements in the vicinity of Γ for samples at three doping levels (x = 0, 1%, 3%, with onset Tc ≈ 18, 16, 15K, correspondingly). As seen in Figs. 2d-2f, the band top shifts to 4 meV and 8 meV below $E_F$ at Co contents of 1% and 3%, respectively. This shift is also clearly demonstrated by the energy distribution curves (EDCs) shown in Figs. 2j-2l. While the low-energy quasiparticle (QP) peaks of pristine LiFeAs are clearly cut off by the Fermi-Dirac (FD) function, those of the 1%Co and 3%Co samples shift below $E_F$ with small spectral weight at $E_F$ due the finite peak-width. We also measured the $k_z$ dispersion of the α band and confirmed its two dimensionality. As an example shown in the supplementary materials[21], the α band of the 3%Co sample is sinking completely below $E_F$ at the Brillouin zone (BZ) centre for all $k_z$ planes. The disappearance of the α FS reduces the density-of-states (DOS) near $E_F$ and hence significantly suppresses the inter-band scattering between the α band and the electron

FSs at the BZ corner, as seen from Figs. 2j-2l.

In the SC state, electrons are gapped toward higher binding energies and form a well-defined Bogoliubov quasiparticle (BQP) peak. Fig. 2j compares the representative EDCs of pristine LiFeAs across the Γ point above and below $T_c$. The electronic states within the SC gap are significantly altered while the states at higher binding energies are only slightly modified by the Bogoliubov dispersion:

$$E_k = \sqrt{\varepsilon_k^2 + \Delta_k^2} \qquad (1)$$

where $E_k$ is the energy of the BQP, which will not show clear deviation from the normal state (NS) energy $\varepsilon_k$, when $\varepsilon_k$ is much larger than the SC gap $\Delta_k$. Interestingly, we find that the electronic states of the 1%Co and 3%Co samples are clearly transferred to higher-binding energies as shown in Figs. 2k and 2l, indicating the opening of a SC gap on the α band, even though this α band is located below $E_F$ at these doping levels.

To see how the SC condensation affects the electronic states away from $E_F$, we show the simulated EDCs without FD distribution in the NS and SC state in Fig. 3b, corresponding to the vertical line at the Γ point in the simulated NS intensity plot in Fig. 3a. The band tops are set at 8 meV and 20 meV below $E_F$, respectively, consistent with the real band positions in this material[21]. With a 5 meV SC gap turned on, the low-energy peak (P1) is shifted by 2 meV while the high-energy peak (P2) is almost unchanged. Moreover, due to the particle-hole mixing, a small peak above $E_F$ develops in the EDC of the SC state. Back to the experimental data, in Figs. 3c, 3e and 3g, the EDCs at the BZ centre exhibit a development of the coherent peak at all the three doping levels below $T_c$. Figs. 3d, 3f and 3h show the same data but divided by the FD function convoluted with the system resolution. The particle-hole mixing in the SC state appears at all doping levels, thus proving the opening of a SC gap. To extract the SC gap, the EDCs in the NS and SC state are fitted by the BCS spectral function plus a constant background[21]. The extracted SC gap keeps almost constant while the line-width of the coherent peak becomes broader at higher doping levels. This indicates that the Co substitution introduces impurity potentials, which is believed to play a destructive role in sign-reversal pairing, at least in the weak coupling regime[1].

Following the procedure shown in Fig. 3, we extract the low-energy band dispersion below and above $T_c$. Figs. 4a and 4b show the extracted data of the 1%Co sample in wide and narrow energy ranges, respectively. In agreement with Eq. 1, the band shift is the largest near the band top and quickly vanishes at higher binding energies. By using the NS data to fit the dispersion in the SC state, we extract the SC gaps of pristine LiFeAs, 1%Co and 3%Co and plot them as a function of the Co concentration in Fig. 4c. The SC gap on the α band remains almost constant while the associated FS topology undergoes a Lifshitz transition with the substitution of Co. Moreover, this SC gap is found to be the largest within the whole BZ space, which rules out the possibility of a proximity effect causing by the pairing on other FS sheets.

As discussed before, the inter-band scattering in the particle-particle channel is dramatically reduced due to the disappearance of the α FS. According to the weak coupling theories, the SC gap on the α band is expected to exponentially decrease when the DOS at $E_F$ goes to zero, as shown in Fig. 4c. This is clearly in contradiction with the experimental observations, which indicates that the SC pairing on the α band is strong and robust. It is also interesting to compare the observed SC gap with $\varepsilon_F$, here defined as the energy difference from the band top to the chemical potential, as illustrated in Fig. 4a. Previous studies on the iron-chalcogenide superconductor FeTe$_{1-x}$Se$_x$[22,23] show that the SC gap is comparable with $\varepsilon_F$ and consistent with a BCS-BEC crossover scenario. In LiFe$_{1-x}$Co$_x$As, as shown in Fig. 4c, the value of $\varepsilon_F$ drops from +2 meV to -8 meV with 3% Co substitution, while the SC gap remains almost unchanged and is immune to the Lifshitz transition. The strong and robust pairing strength with negative $\varepsilon_F$ is in sharp contrast to the weak coupling BCS-like theory, where superconductivity is treated as a perturbation theory based on the assumption that $\Delta_{SC} \ll \varepsilon_F$.

Recent inelastic neutron scattering (INS) and nuclear magnetic resonance (NMR) studies have demonstrated that low-energy spin fluctuations are relatively weak and incommensurate for pristine LiFeAs[24-26]. Combined with ARPES data, the incommensurate INS peaks were attributed to the inter-band scattering between the β FS and the electron FSs[27]. However, ARPES and STM studies demonstrate that the largest SC gap is indeed on the α band, which is clearly incompatible with the low-energy spin fluctuations scenario[19,20,28,29]. Although this discrepancy can be removed by subtle modifications involving orbital fluctuations or small-q inter-band scattering[30,31], our observation of strong SC pairing on the bands without FS here is beyond any reasonable mending within the weak coupling approach. Instead, it is naturally consistent with many strong coupling approaches[6-10,32], in particular with the $J_1$-$J_2$ model that predicts the strongest pairing at the zone centre[6-9].

In summary, we have observed an unprecedented strong pairing on energy band without FS at the BZ centre of $LiFe_{1-x}Co_xAs$. The observed SC pairing strength is strong and robust against the reduction of the DOS near $E_F$ and the increasing impurity scatterings. The immunity of the SC pairing across the Lifshitz transition rules out the fundamental assumption of weak coupling theories that superconductivity is a FS instability. Our results clearly demonstrate that the pairing mechanism of the IBSCs resides in the strong coupling regime.

**Method:**

Single crystals of $LiFe_{1-x}Co_xAs$ were synthesized by the self-flux method using $Li_3As$, $Fe_{1-x}Co_xAs$ and As powders as the starting materials. The $Li_3As$, $Fe_{1-x}Co_xAs$ and As powders were weighed according to the element ratio of $Li(Fe_{1-x}Co_x)_{0.3}As$. The mixture was grounded and put into an alumina crucible and sealed in Nb crucibles under 1 atm of Argon gas. The Nb crucible was then sealed in an evacuated quartz tube, heated to 1100 °C and slowly cooled down to 700 °C at a rate of 3 °C/hr. High energy resolution ARPES data were recorded at the Institute of Physics, Chinese Academy of Sciences, using the He Iα ($h\nu$ = 21.218 eV) resonance line of a helium discharge lamp. The angular and momentum resolutions were set to 0.2° and 3 meV, respectively. ARPES polarization-dependent measurements were performed at Beamline I05 of Diamond Light Source using a Scienta R4000 with energy and momentum resolutions set to 0.2° and 10 meV, respectively. To select the α band, we employed linearly polarized light with the potential vector perpendicular to the mirror plane of the sample. All samples were cleaved *in situ*. The data was taken in a vacuum better than $3*10^{-11}$ Torr with discharge lamp and $1*10^{-10}$ Torr with synchrotron light source.

**Acknowledgement:**

We thank A. Chubukov, Y. M. Dai, W. Ku，X. R. Liu for useful discussions. This work was supported by grants from CAS (2010Y1JB6 and XDB07000000), MOST (2010CB923000, 2011CBA001000 and 2013CB921700), NSFC (11234014 and 11274362).

**Contribution:**

H. M. and H. D. designed the experiments, H. M., T. Q., X. S., P. R., T. K. and M. H. carried out the experiments; H. M. analysed the data; L.-Y. X., X.-C. W. and C.-Q. J. provided the samples; H. M., T. Q., P. R., J.-P. H. and H. D. wrote the manuscript. All authors discussed the results and commented on the manuscript.

**Figure 1 | Fermi surface topology of pristine LiFeAs and LiFe$_{0.97}$Co$_{0.03}$As. a** and **b,** Plots of the ARPES intensity at $E_F$ of LiFeAs ($T_c$ = 18 K) and LiFe$_{0.97}$Co$_{0.03}$As ($T_c$ = 15 K) as a function of the two-dimensional wave vector measured at 30 K with the He Iα line ($h\nu$ = 21.218 eV). The intensity is obtained by integrating the spectra within 10 meV with respect to $E_F$. **c** and **d**, Extracted $k_F$ loci of LiFeAs and LiFe$_{0.97}$Co$_{0.03}$As, respectively. The small hole-like FS at the BZ centre is sinking below $E_F$ due to the substitution of Co, and expected to significantly suppress the inter-band scattering between the electron and hole FSs.

**Figure 2 | Band dispersions of LiFe$_{1-x}$Co$_x$As near $E_F$. a-c,** ARPES intensity plots of pristine LiFeAs, 1%Co and 3%Co samples, respectively, in the NS across the BZ centre. The data are recorded with the He Iα line ($h\nu$ = 21.218 eV), which is close to $k_z$ = 0. **d-f,** Same data as **a-c**, but divided by the Fermi-Dirac function convoluted with the system resolution. Red circles are the extracted NS dispersion of the α band. The α band is sinking below $E_F$ in the 1%Co and 3%Co compounds. **g-i,** Corresponding intensity plots in the SC state. The difference between the NS and the SC state is clearly resolved in **j-l,** where representative EDCs in both the NS and the SC state are plotted together. The enhanced spectral weight at the low-binding energy is attributed to the coherence of the paired electrons.

**Figure 3 | Extracting the SC gap from EDCs. a,** ARPES intensity plot that simulates the NS band dispersion without FD distribution of 3%Co sample. **b,** Simulated EDCs in the NS (red) and SC state (blue), corresponding to the vertical line in the intensity plot in **a**. The spectral function is assumed to have the BCS form[21]. The small peak above $E_F$ in the blue EDC is due to the particle-hole mixing, which is a hallmark of SC condensation. By using the BCS spectral function to fit the EDCs in the NS and the SC state, we extract the SC gap. **c**, **e** and **g**, EDCs at the BZ centre in the NS and the SC state of pristine LiFeAs, 1%Co and 3%Co samples, respectively. To extract the SC gap, the raw data shown in **c, e** and **g** are divided by FD function convoluted with the system resolution and shown in **d**, **f**, and **h**, respectively. The particle-hole mixing in the SC state appears at all doping levels, thus proving the opening of SC gap. Green dotted curves on light blue and red circles are the fitting results. The decomposed spectral functions in the SC state are appended below the fittings. The broadening of the decomposed peaks at higher doping levels is likely caused by an enhanced impurity scattering due to the in-plane substitution.

**Figure 4 | Extracting the SC gap from dispersions and robustness of the pairing against disappearance of FS. a**, Extracted band dispersions of LiFe$_{0.99}$Co$_{0.01}$As below (light blue squares) and above (pink circles) $T_c$. A zoom in to a narrower energy range in **b** shows that the difference between the NS and the SC state is the largest at the BZ centre, and then gets smaller at high binding energies. The green dotted curve is the fitting result, which yields a similar SC gap as extracted from EDCs. **c**, Comparison of the extracted SC gap from the experimental data and the expected ones from weak coupling theories. This shows that the SC gap observed on the α band is robust despite the disappearance of the α FS at the Γ point. This observation is incompatible with the weak coupling theory, in which the SC gap is expected to be sensitive to both the DOS near $E_F$ and impurity scattering.

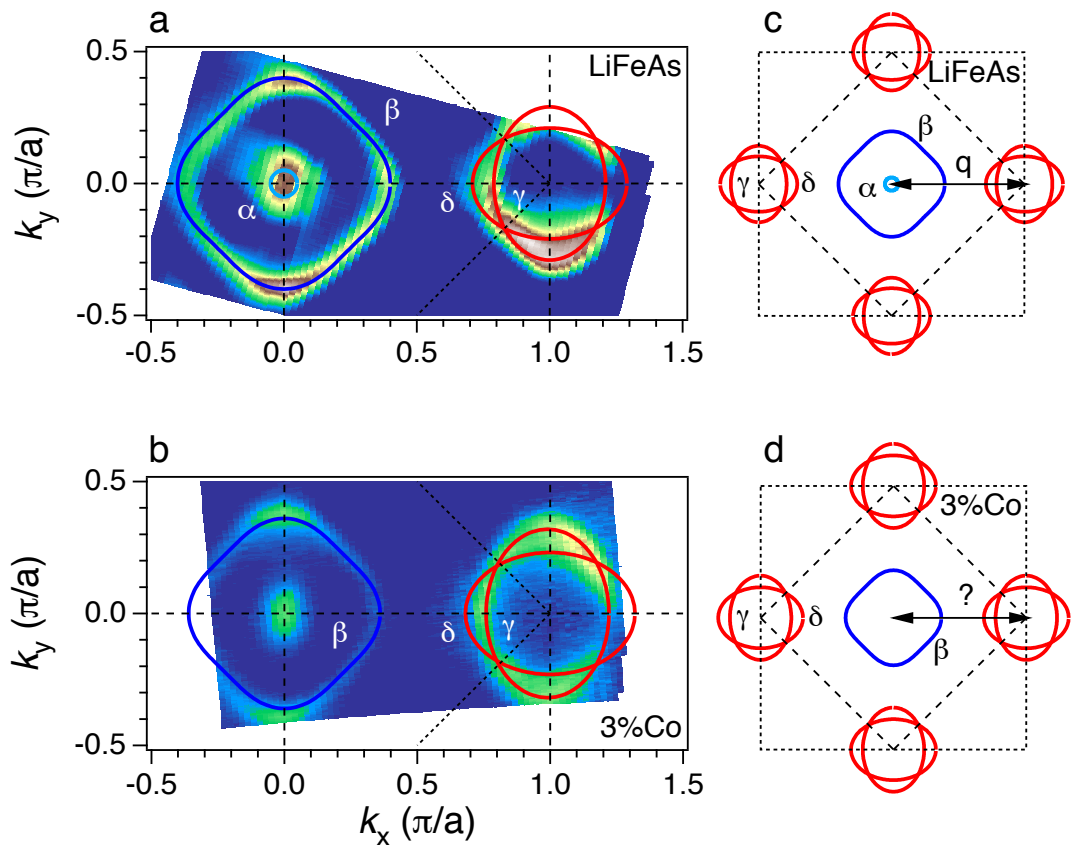

Figure 1

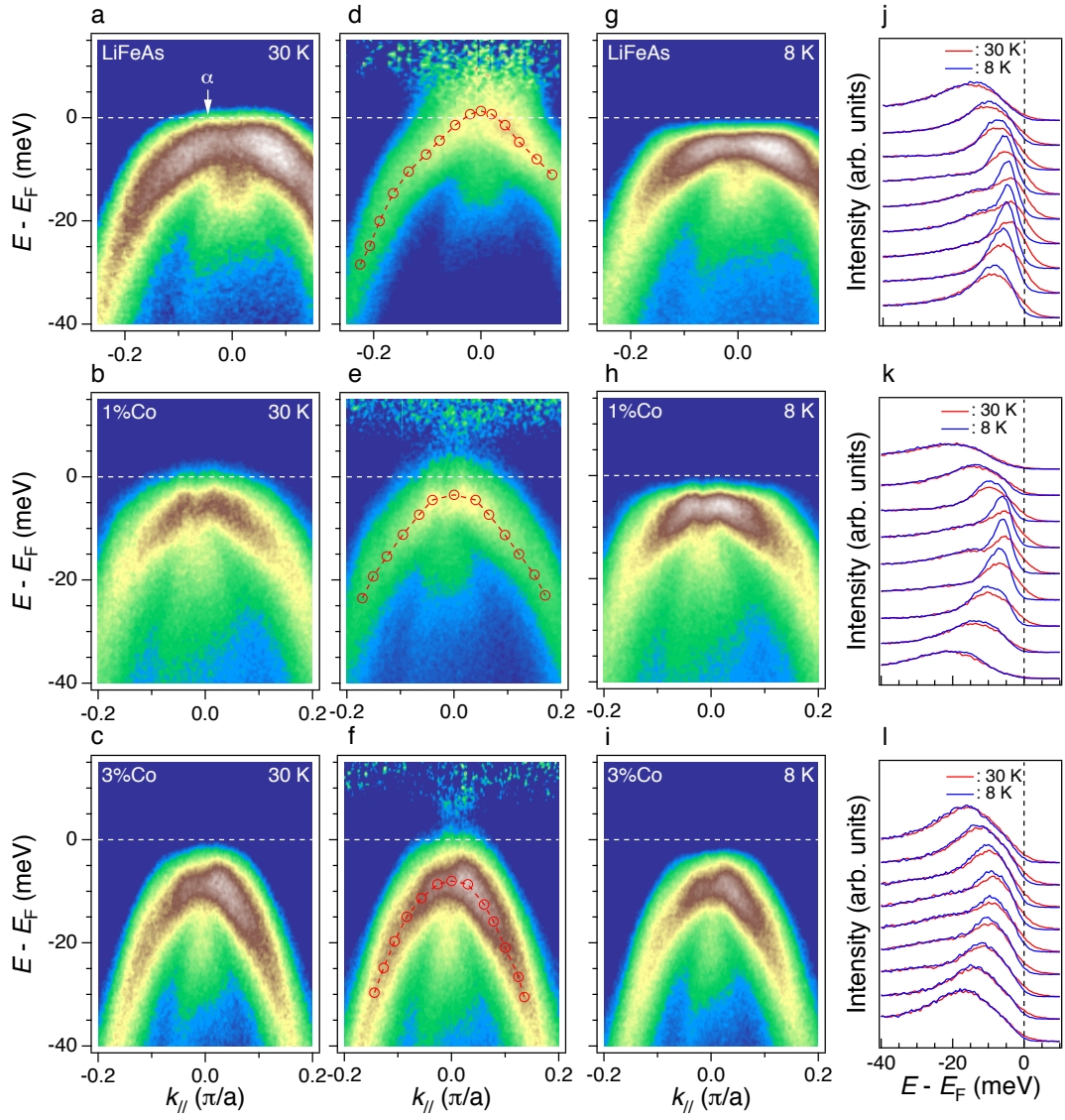

Figure 2

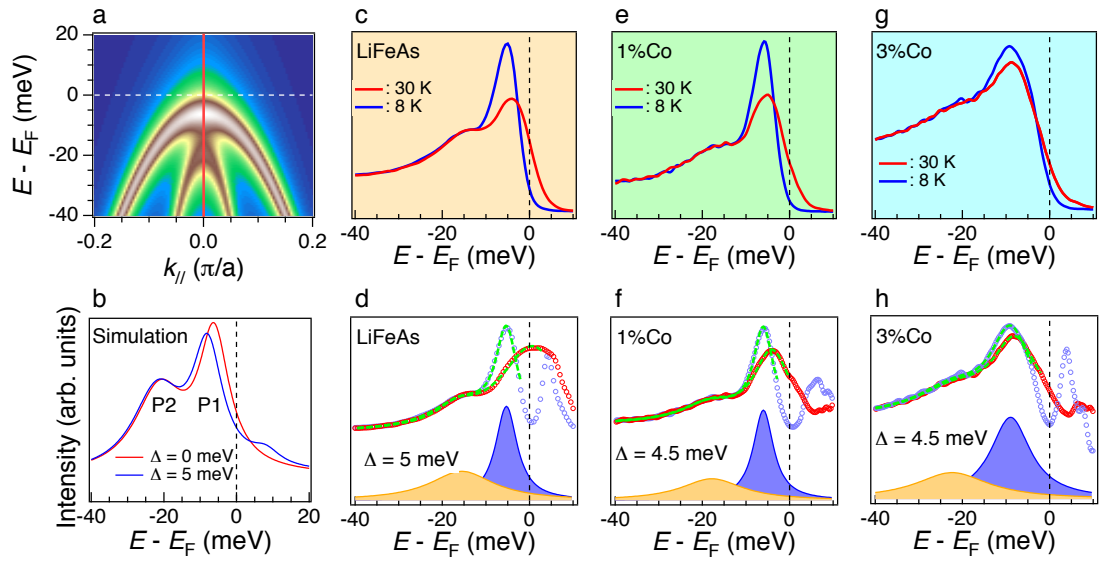

Figure 3

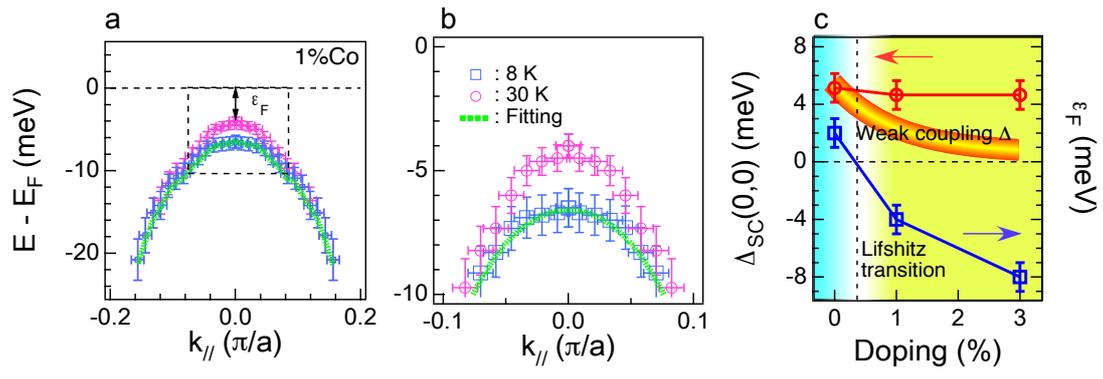

Figure 4

# Supplementary materials for

Observation of strong electron pairing on bands without Fermi surfaces in LiFe$_{1-x}$Co$_x$As

H. Miao[1], T. Qian[1,*], X. Shi[1], P. Richard[1,2], T. K. Kim[3], M. Hoesch[3], L. Y. Xing[1], X. –C. Wang[1], C. –Q. Jin[1,2], J. –P. Hu[1,2,4] and H. Ding[1,2,*]

To whom correspondence should be addressed.

*: tqian@iphy.ac.cn

‡: dingh@iphy.ac.cn

This PDF file includes:

Supplementary text

Figs. S1 to S5

References

**Table of contents:**

## 1. Transport properties of LiFe$_{1-x}$Co$_x$As (x = 0, 0.01, 0.03)

Figs. S1a and S1b show the resistivity and the magnetic susceptibility of LiFe$_{1-x}$Co$_x$As. The low residual resistivity and high superconducting (SC) volume prove the high quality of our samples.

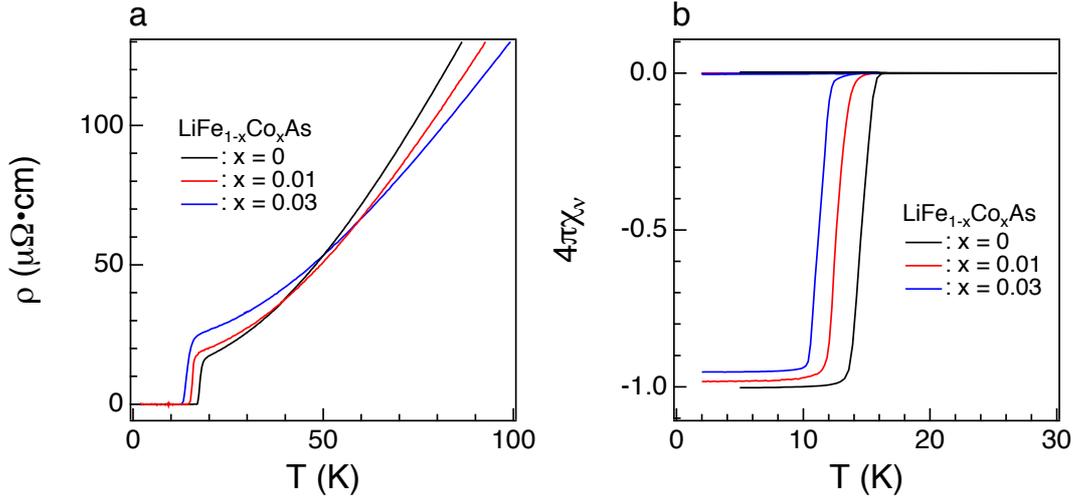

**Figure S.1: Resistivity and magnetic susceptibility of LiFe$_{1-x}$Co$_x$As**

## 2. FS topology of LiFe$_{0.99}$Co$_{0.01}$As

Fig. S2a shows the ARPES intensity at $E_F$ of LiFe$_{0.99}$Co$_{0.01}$As as a function of the two-dimensional wave vector measured with the He I$\alpha$ line ($h\nu$ = 21.218 eV). The intensity is obtained by integrating the spectra within 10 meV with respect to $E_F$ and the energy resolution is set to 14 meV. To directly compare the FSs evolution as a function of doping, we summarized the extracted FSs in Fig. S2b. Black, red and blue curves represent the pristine LiFeAs, Co_1% and Co_3% samples, respectively. Our results confirm that the substitution of Co introduces electron carriers and reduces/expands hole/electron FSs[1].

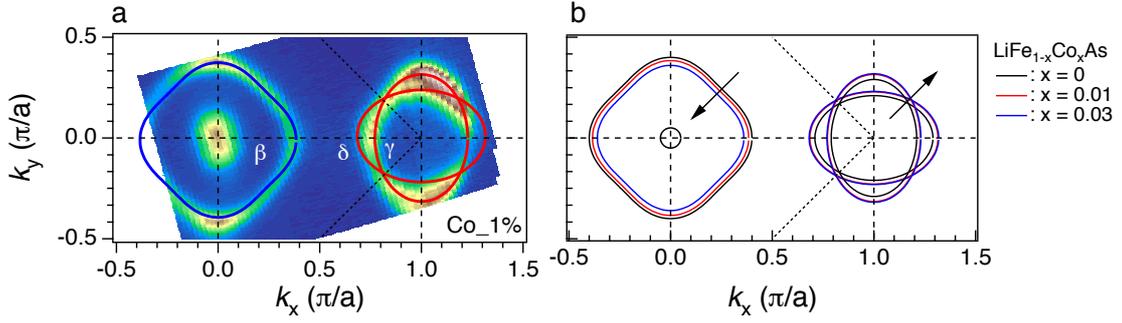

**Figure S.2: FS evolution of LiFe$_{1-x}$Co$_x$As**

## 3. The $k_z$ effect

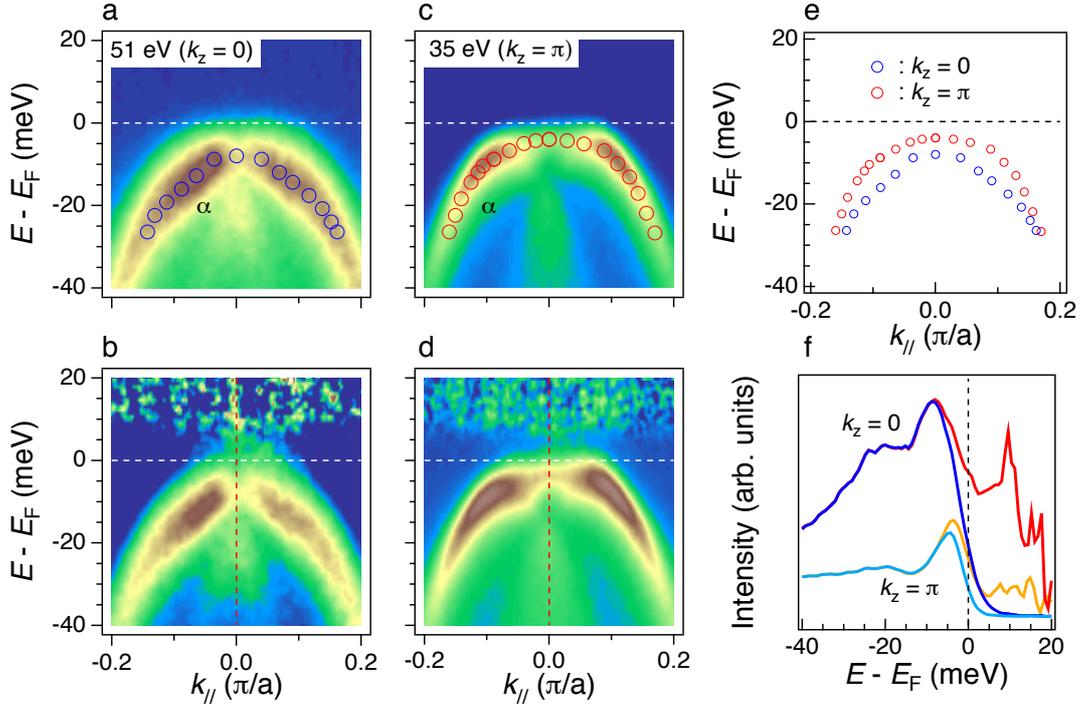

**Figure S.3**: **Band dispersion of the α band at $k_z$ = 0 and π in LiFe$_{0.97}$Co$_{0.03}$As**

Although the electronic structure of LiFeAs is quite two-dimensional, previous studies indicate that the α band has a small dispersion along $k_z$[1,2,3]. To quantify the $k_z$ effect of the α band in LiFe$_{0.97}$Co$_{0.03}$As, we extract the normal state (NS) dispersion of the α band at $k_z = 0$ and π. Figs. S3a and S3c show the ARPES intensity plots of LiFe$_{0.97}$Co$_{0.03}$As at 51 eV ($k_z = 0$) and 35 eV ($k_z = π$), respectively. The data are recorded at 20 K with linearly polarized light to select the α band. Blue and red circles on top of the intensity are extracted from EMDCs. Figs. S3b and S3d are the same data but divided by the Fermi-Dirac function convoluted with the system resolution. Fig. S3f shows the EDCs at the Brillouin zone centre as marked by the red dashed line in Fig. S3b and Fig. S3d. The intensity plots and EDCs show that the band top of the α band has tiny dispersion along $k_z$. Our results prove that at least for LiFe$_{0.97}$Co$_{0.03}$As, the α band is completely sinking below $E_F$ all over the momentum space.

## 4. Extraction of the SC gap without FS

To extract the SC gap from EDCs, we use functions:

$$I(\omega) = C_0 + \frac{a_1}{\left((\omega - E_k^1)^2 + \Gamma_1^2\right)} + \frac{a_2}{\left((\omega - E_k^2)^2 + \Gamma_2^2\right)} \quad (1)$$

$$E_k^i = \sqrt{\left((\varepsilon_k^i)^2 + \Delta_k^2\right)} \quad (2)$$

$$a_i \propto \Gamma_i (u_k^i)^2 = \frac{1}{2} \Gamma_i \left(1 + \frac{\varepsilon_k^i}{E_k^i}\right) \quad (3)$$

Here we assume a BCS spectral function in the occupied states, with the $\Gamma_k$ parameter not changing with binding energy. $a_i$ is a fitting constant, which is proportional to $\Gamma_k$ and $u_k^2$. The fitted results and extracted SC gap are shown in Fig. 3 of the main text.

## 5. Extracted band dispersions of LiFe$_{0.97}$Co$_{0.03}$As

In Fig. 4 of the main text, we show the dispersion of the α band in LiFe$_{0.99}$Co$_{0.01}$As and use the NS and SC state dispersions to extract the SC gap. We used similar procedure with LiFe$_{0.97}$Co$_{0.03}$As. Figs. S4a and S4b show the extracted and fitted results. Pink circles and blue squares are the extracted NS and SC state dispersions of the α band, respectively. The green dashed curve is obtained by using the NS dispersion to fit the SC state dispersion.

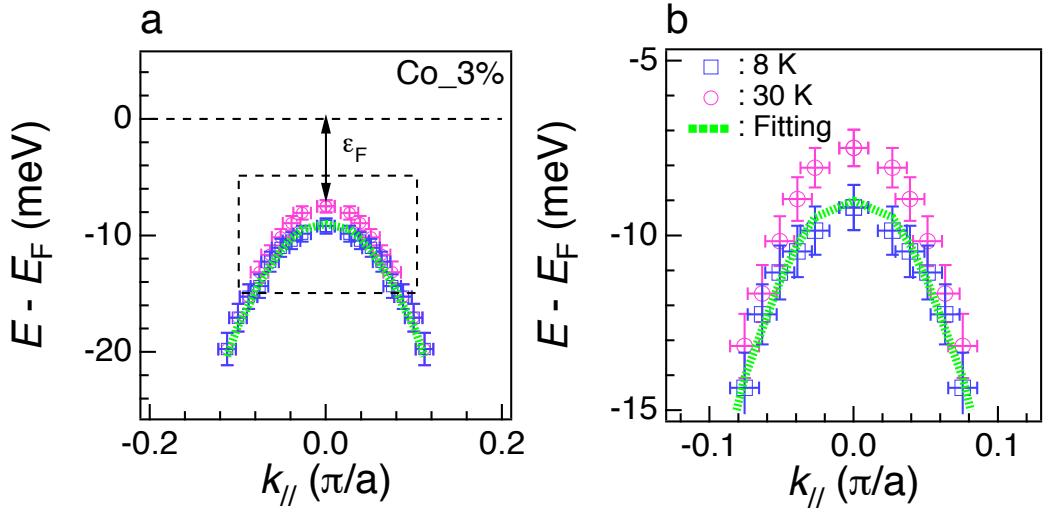

**Figure S.4**: Extracted NS and SC state dispersions of the α band in LiFe$_{0.97}$Co$_{0.03}$As

## 6. The superconducting gap on the β and electron FSs.

As shown in the main text, the SC gap on the α band is large and robust.

It is important to show how the SC gaps on other FSs change with the substitution of Fe by Co. Fig. S5 shows the SC gap measured on the β and electron FSs. The measured $k_F$ positions are illustrated in Figs. S5a and S5e. ARPES intensities shown in Figs. S5b, S5c S5f and S5g are divided by the FD function convoluted with the system resolution. As shown in Figs. S5d and S5h, the SC gap on β, δ and γ FSs are 3 meV, 3.5 meV and 4.2 meV respectively. Our results prove that the SC gap on the α band is the largest over the momentum space and thus rule out a proximity effect origin.

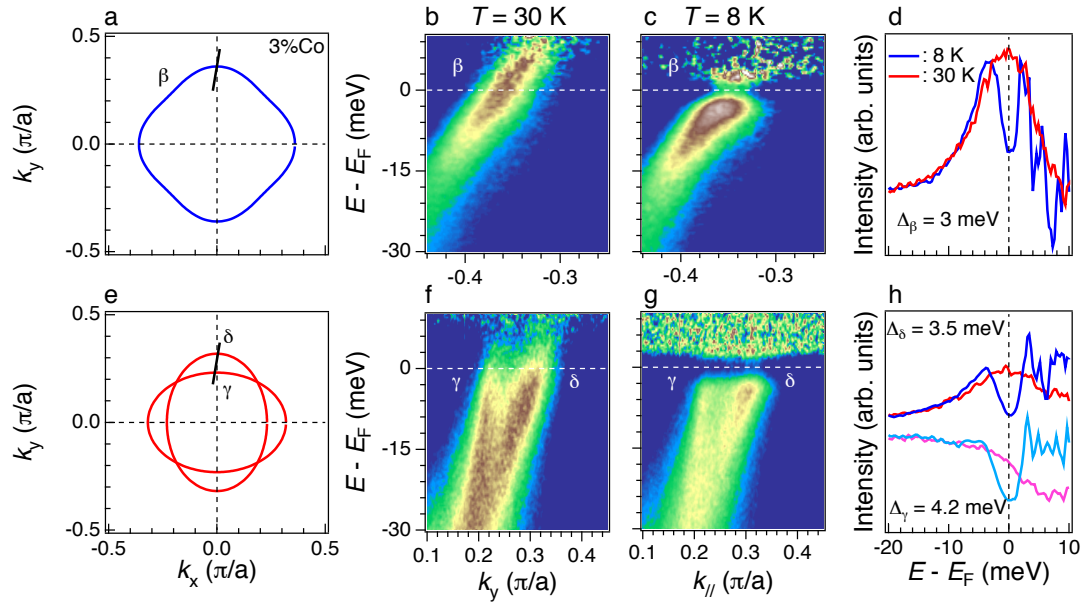

**Figure S.5: The superconducting gap on β and electron FSs**